\documentclass[12pt]{iopart}

\usepackage{iopams}
\usepackage{graphicx}
\usepackage{color}
\usepackage{cite}
\usepackage[normalem]{ulem}

\newcommand{\ctrip}{${\rm c}^3\Sigma^+$~}
\newcommand{\Bsing}{${\rm B}^1\Pi$~}

\newcommand{\Bcres}{${\rm B}^1\Pi{\sim}{\rm c}^3\Sigma^+$~}
\newcommand{\nak}{$^{23}$Na$^{40}$K}
\newcommand{\cm}{$\rm cm^{-1}$}

\begin{document}

\title[]{Two-Photon Pathway to Ultracold Ground State Molecules of $^{23}$Na$^{40}$K
}

\author{Jee Woo Park, Sebastian A.~Will, and Martin W.~Zwierlein\footnote{Corresponding author: zwierlei@mit.edu}}

\address{ Massachusetts Institute of Technology, 77 Massachusetts Avenue, Cambridge, MA 02139, USA}



\begin{abstract}
We report on high-resolution spectroscopy of ultracold fermionic \nak~Feshbach molecules, and identify a two-photon pathway to the rovibrational singlet ground state via a resonantly mixed \Bcres intermediate state. Photoassociation in a $^{23}$Na-$^{40}$K atomic mixture and one-photon spectroscopy on \nak~Feshbach molecules reveal about 20 vibrational levels of the electronically excited \ctrip state. Two of these levels are found to be strongly perturbed by nearby \Bsing states via spin-orbit coupling, resulting in additional lines of dominant singlet character in the perturbed complex {${\rm B}^1\Pi |v{=}4\rangle {\sim} {\rm c}^3\Sigma^+ | v{=}25\rangle$}, or of resonantly mixed character in {${\rm B}^1\Pi | v{=}12 \rangle {\sim}{\rm c}^3\Sigma^+ | v{=}35 \rangle$}. The dominantly singlet level is used to locate the absolute rovibrational singlet ground state ${\rm X}^1\Sigma^+ | v{=}0, J{=}0 \rangle$ via Autler-Townes spectroscopy.
We demonstrate coherent two-photon coupling via dark state spectroscopy between the predominantly triplet Feshbach molecular state and the singlet ground state. Its binding energy is measured to be 5212.0447(1) \cm, a thousand-fold improvement in accuracy compared to previous determinations. In their absolute singlet ground state, \nak~molecules are chemically stable under binary collisions and possess a large electric dipole moment of $2.72$ Debye. Our work thus paves the way towards the creation of strongly dipolar Fermi gases of NaK molecules.
\end{abstract}

\vspace{2pc}

\maketitle

\section{Introduction}

Ever since laser cooling and evaporative cooling gave full control over the motional and internal degrees of atoms, there has been a strong effort to extend such control over the richer internal structure of molecules. With large samples of molecules, all occupying the same internal quantum state, controlled switching of chemical reactions could be studied at the quantum level. Their internal degrees of freedom make molecules suitable as potential carriers for quantum information, as a quantum resource for precision measurements and as the building block for novel quantum many-body systems. A particular appeal lies in the creation of new forms of quantum dipolar matter, such as topological superfluids or quantum crystals.
The enormous progress towards ultracold molecules over the last decade has been summarized in review papers~\cite{Carr09mol, Baranov2012, Quemener:2012}, to which one may add the recent successes of magneto-optical trapping~\cite{Hummon2013MoleculeMOT,Barry2014MoleculeMOT}, a novel type of Sisyphus cooling~\cite{Zeppenfeld2012Sysiphus}, as well as evaporative cooling of molecules~\cite{Stuhl2012Evapcoolingmolecules}, among others. An alternative to direct cooling of molecules, which has so far been limited to temperatures of several millikelvin, is the creation of weakly bound Feshbach molecules at nanokelvin temperatures from a gas of ultracold atoms, which has been achieved for homonuclear~\cite{rega03mol,cubi03,stre03,joch03lith,chin03feshbach_mol,xu03na_mol,durr04mol} as well as for heteronuclear molecules~\cite{stan04,inou04,ospe06hetero,Spiegel2010LiK,Takekoshi2012RbCs,wu2012NaK,Heo2012NaLi,Koppinger2014RbCs}. Feshbach molecules created out of fermionic atoms were remarkably long-lived~\cite{cubi03}, allowing the observation of Bose-Einstein condensation~\cite{grei03molbec,joch03bec,zwie03molBEC,bour04coll,bart04, part05}. However, the dipole moment of such long-range Feshbach molecules is still vanishingly small. In pioneering works at JILA and in Innsbruck, ultracold molecules of fermionic $^{40}$K$^{87}$Rb and bosonic non-polar $^{133}\rm Cs_2$ in their absolute rovibrational ground state were created from weakly bound Feshbach molecules via a coherent two-photon transfer~\cite{danzl08Csmol,Danzl2010groundstate,ni08polar}. $^{40}$K$^{87}$Rb molecules were found to undergo the reaction $2 \rm KRb \rightarrow K_2 + Rb_2$~\cite{ospe10chemical}, precluding further cooling into the deeply degenerate regime. However, confinement to two- and three-dimensional optical lattice potentials allowed to stabilize molecular gases of KRb against reactive collisions \cite{Chotia:2012,Yan2013dipolarspin}. Recently, ultracold gases of bosonic RbCs have been created~\cite{Takekoshi2014RbCs,Molony2014RbCs}, a molecule that is  chemically stable under binary collisions~\cite{zuch10mol}. 

In the present work we focus on the fermionic molecule \nak, also chemically stable under two-body collisions~\cite{zuch10mol} and known to possess a large electric dipole moment of 2.72 Debye~\cite{worm81nak}, about five times larger than that of KRb. It is therefore an ideal candidate for the formation of a strongly interacting, stable dipolar Fermi gas with interaction energies on the order of tens of percent of the Fermi energy.
Following our creation of ultracold Feshbach molecules~\cite{wu2012NaK}, we here perform one- and two-photon spectroscopy on \nak~and identify a two-photon pathway from the predominantly triplet Feshbach molecular state to the absolute rovibrational singlet ground state. A priori, the existence of such a pathway is not obvious, as spin-orbit coupling, mixing singlet and triplet states, is weak for this light molecule.

Historically, NaK was a textbook example of a diatomic molecule and one of the best studied diatomic molecules before the advent of photoassociation and coherent two-photon spectroscopy of cold and ultracold KRb. The earliest absorption spectra~\cite{Barratt1924NaK} predate modern quantum mechanics, but already within a decade the singlet ground and first excited Born-Oppenheimer potentials had been experimentally characterized~\cite{Loomis1934NaK}. In particular, levels of the ${\rm B}^1\Pi$ state were already known to an accuracy of 1~cm$^{-1}$. In the era of laser spectroscopy, this accuracy improved to 0.001~cm$^{-1}$, resulting in highly precise excited singlet potentials, for example of the ${\rm B}^1\Pi$ state relevant here~\cite{Kato1990c3s,Kasahara1991B1Pi}, and of the singlet and triplet ground states, ${\rm X}^1\Sigma^+$ and ${\rm a}^3\Sigma^+$~\cite{Russier2000NaK,gerd08nak}. In addition, singlet-triplet perturbations were observed in electronically excited states that led to fluorescence down to the triplet ground state, first ${\rm D}^1\Pi\rightarrow {\rm a}^3\Sigma^+$~\cite{Breford1978NaK,Breford1979a3s,Eisel1979NaK,Kato1980NaK}, later ${\rm B}^1\Pi\rightarrow {\rm a}^3\Sigma^+$~\cite{Breford1981NaKperturbation,Barrow1987B1Pi,Baba1988NaK,Kowal1989NaKtriplet,Kato1990c3s,Ishikawa1992c3shyperfine}. This already demonstrates that two-photon coupling is possible between the triplet and singlet ground states, and indeed, one of the perturbations studied here (albeit in the previously unstudied isotopologue \nak) was first observed in laser-induced fluorescence in~\cite{Breford1981NaKperturbation}. The excited triplet states that caused such perturbations were characterized via deperturbation techniques~\cite{Field2004spectra}, leading to deperturbed potential energy curves for the ${\rm b}^3\Pi$ state~\cite{Ross1986b3Pi} and the ${\rm c}^3\Sigma^+$ state~\cite{Kato1990c3s,Kowal1995c3s,Ferber2000NaK}. These potentials and calculated energy levels are used in the present study as a guide for our spectroscopic search for perturbations in \nak, which has not been studied before due to $^{40}$K's low natural abundance of 0.01\%. Recently, two-photon spectroscopy in the ``traditional direction'' from the absolute singlet to low-lying triplet ground states of NaK was performed on $^{23}$Na$^{39}$K in a molecular beam~\cite{Schulze:2013}.

\begin{figure}
  \begin{center}
  \includegraphics[width=0.6\columnwidth]{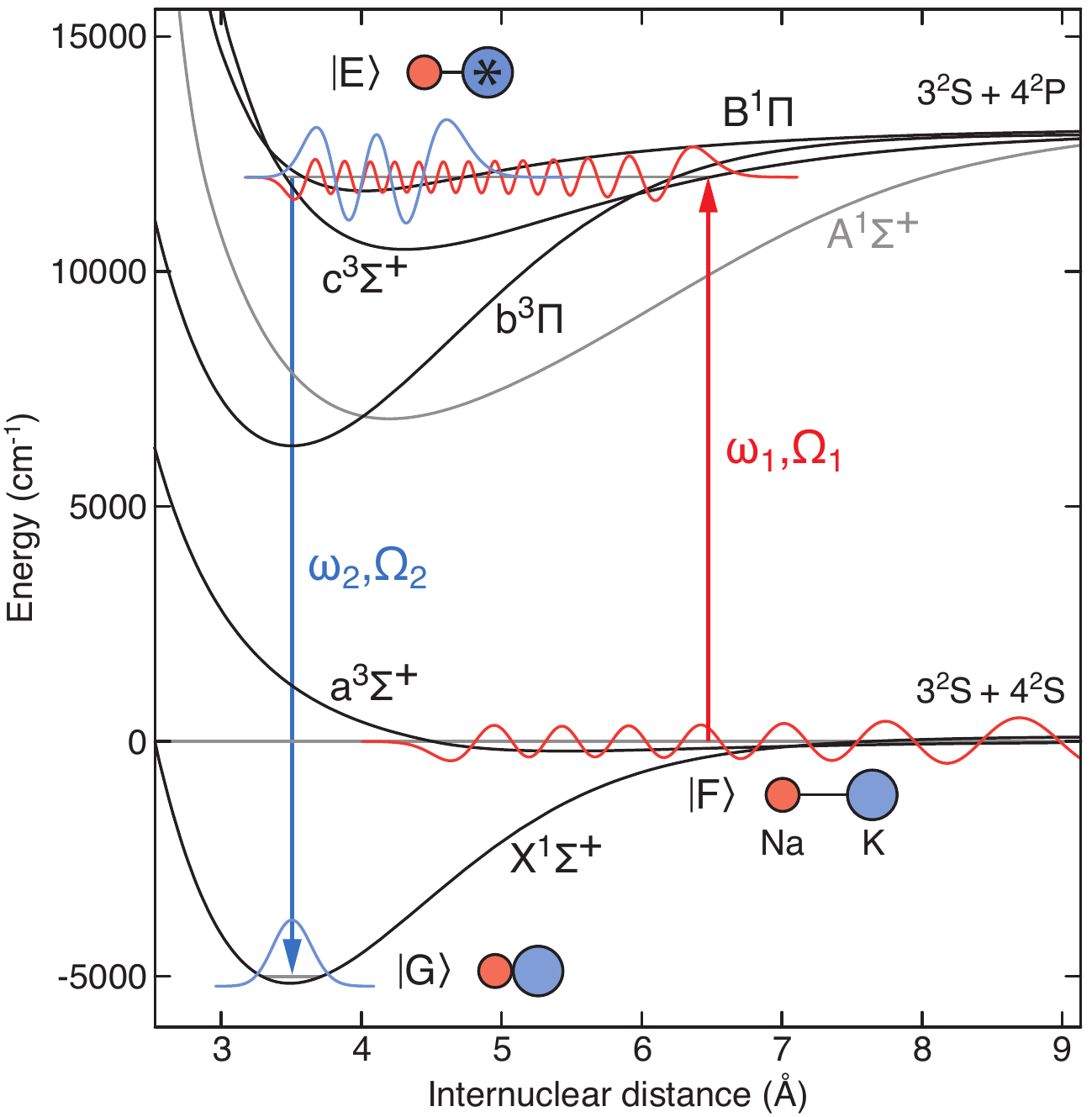}
  \end{center}
  \caption{\label{fig:cartoon} Ground and excited state molecular potentials of NaK relevant to the two-photon STIRAP transfer to the absolute rovibrational ground state $|\rm G\rangle$ in the ${\rm X}^{1}\Sigma^+$ potential. The initial state of weakly bound Feshbach molecules $|\rm F\rangle$ is associated with the highest vibrational level $|v{=}19\rangle$ of the ${\rm a}^{3}\Sigma^+$ potential. Excited states $|{\rm E}\rangle$ explored in this work are dominantly associated with the ${\rm c}^{3}\Sigma^+$ and ${\rm B}^{1}\Pi$ potentials, and also influenced by states of the ${\rm b}^{3}\Pi$ potential. As an example, the vibrational wavefunctions of the coupled complex ${\rm B}^{1}\Pi| v{=}4\rangle{\sim} {\rm c}^{3}\Sigma^+| v{=}25\rangle$ are shown. The states of the three-level system $|{\rm F}\rangle$, $|{\rm E}\rangle$ and $|{\rm G}\rangle$ are coupled by up-leg and down-leg lasers with frequencies $\omega_1$, $\omega_2$ and Rabi couplings $\Omega_1$, $\Omega_2$.}
\end{figure}

We here demonstrate that even the highest vibrationally excited triplet ground state near a Feshbach resonance can be coherently coupled to the absolute rovibrational singlet ground state with significant coupling strength, enabling the formation of ground state dipolar molecules from ultracold Feshbach molecules. The potential energy curves are shown in Fig.~\ref{fig:cartoon}. The explored pathway connects the predominantly triplet Feshbach molecular state to a perturbed level of mixed \Bcres character via the up-leg laser of frequency $\omega_1$. From there, the down-leg laser of frequency $\omega_2$ promotes the coupling to the absolute singlet ground state.

The starting point of the experiment is a near degenerate Bose-Fermi mixture of $^{23}$Na-$^{40}$K, as described in our previous work~\cite{Park:2012}. Since in heatpipe or beam experiments only the deep singlet molecular levels are populated, the triplet states ${\rm c}^3\Sigma^+$ had only been observed where they were strongly perturbed by a nearby ${\rm B}^1\Pi$ state. There are thus uncertainties in the exact location of ${\rm c}^3\Sigma^+$ energy levels of more than 1~cm$^{-1}$. The first task is therefore to perform photoassociation spectroscopy on the atomic ultracold Na-K mixture, as described in section~\ref{s:PA}. This allows the direct observation of ${\rm c}^3\Sigma^+$ levels, and thanks to the ultracold temperatures, only low lying rotational states are accessed.
Equipped with the exact locations of the ${\rm c}^3\Sigma^+$ levels, and highly accurate predictions on the locations of singlet ${\rm B}^1\Pi$ levels from mass-scaled Dunham coefficients for $^{23}$Na$^{39}$K~\cite{Kato1990c3s,Kasahara1991B1Pi}, we identify two accidental degeneracies between ${\rm c}^{3}\Sigma^+$ and ${\rm B}^1\Pi$ in \nak, namely {${\rm B}^1\Pi | v{=}4 \rangle {\sim} {\rm c}^3\Sigma^+ | v{=}25 \rangle$} and {${\rm B}^1\Pi | v{=}12 \rangle {\sim}{\rm c}^3\Sigma^+ | v{=}35 \rangle$}, where $v$ denotes the vibrational quantum number. In section~\ref{s:mixing} we therefore perform high-resolution spectroscopy on these near-degenerate levels starting with Feshbach molecules. We resolve detailed fine- and hyperfine structure that is well reproduced by our theoretical calculations. The manifold {${\rm B}^1\Pi | v{=}4 \rangle {\sim} {\rm c}^3\Sigma^+ | v{=}25 \rangle $} is sufficiently perturbed so that the singlet-dominant feature can be directly addressed from triplet Feshbach molecules. For {${\rm B}^1\Pi |v{=}12 \rangle {\sim}{\rm c}^3\Sigma^+| v{=}35 \rangle$}, the observation of two pairs of lines with nearly identical hyperfine structure signals exceptionally strong mixing, originating from an essentially perfect degeneracy and sufficiently strong spin-orbit coupling. Finally in section~\ref{s:groundstate}, we perform two-photon spectroscopy, first to locate the absolute rovibrational singlet ground state using Autler-Townes spectroscopy, then to perform coherent dark state spectroscopy for a high-precision measurement of the binding energy of the rovibrational ground state. The strong singlet-triplet mixing in the intermediate state will enable stimulated rapid adiabatic passage (STIRAP) from Feshbach molecules to absolute singlet ground state molecules. Our work thus opens the door to the creation of a strongly interacting dipolar Fermi gas of \nak~molecules.

\section{Photoassociation spectroscopy of $^{23}$Na - $^{40}$K}
\label{s:PA}

Photoassociation (PA) spectroscopy is commonly applied to laser-cooled atomic gases trapped in magneto-optical traps (MOTs). However, atomic densities in MOTs are limited to about $10^{11} \,\rm cm^{-3}$, and free-bound transitions to excited molecular levels can typically only reach the long-range part of the excited state potentials. PA in a MOT has additional complications for heteronuclear molecules: The excited state is a strong $C_6$ potential, which has a shorter range than the $C_3$ potential in homonuclear molecules; and overlap of the two MOTs results in severe light-assisted collisions, strongly reducing the effective pair density. Instead of using a MOT, we therefore opted for creating a dense Bose-Fermi mixture of $^{23}$Na-$^{40}$K in an optical trap, cooled close to quantum degeneracy. This allows us to work with atomic densities of up to $10^{12} \,\rm cm^{-3}$. The PA laser can illuminate the atomic sample for 10 seconds at high laser power, revealing even deeply bound lines of the excited state potentials.

To summarize the sequence, as outlined in references~\cite{Park:2012,wu2012NaK}, laser cooled $^{23}$Na and $^{40}$K atoms are first captured in a magneto-optical trap and optically pumped to the $|f_{\rm Na} ,m_{f_{\rm Na}}\rangle = | 2, 2 \rangle$ and $|f_{\rm K} ,m_{f_{\rm K}}\rangle = | 9/2, 9/2 \rangle$ states before being transferred into an optically plugged magnetic quadrupole trap. In the magnetic trap, the mixture is cooled by forced radiofrequency (rf) evaporation of $^{23}$Na on the  $|2, 2\rangle\rightarrow |1,1\rangle$ transition. $^{40}$K atoms are sympathetically cooled via collisions with $^{23}$Na atoms. At a temperature of about $2\,\rm\mu$K, the mixture is loaded into a crossed optical dipole trap operating at 1064 nm and each species is transferred into the lowest hyperfine states $|1,1\rangle$ for Na and $|9/2,-9/2\rangle$ for K via Landau-Zener rf sweeps. For the measurements presented in this work, further evaporation in the dipole trap lowers the temperature to about $1\,\rm\mu$K, creating an equal mixture of about $10^5$ atoms of both species close to quantum degeneracy.

PA spectroscopy is performed on the atomic $^{23}$Na-$^{40}$K mixture to explore the electronically excited potentials of  $^{23}$Na$^{40}$K  (see Fig.~\ref{fig:cartoon}). To this end, the near degenerate mixture of Na and K is illuminated with an intense laser beam (laser 1), generated by a tunable continuous wave titanium sapphire laser. The typical peak intensity at the position of the atoms is 7 kW/cm$^2$. The polarization of the light is fixed to be diagonal with respect to the vertical magnetic bias field to address both $\pi$ and $\sigma^{+/-}$ transitions to excited states of $^{23}$Na$^{40}$K. In order to efficiently cover a large spectral range during a single experimental run, the frequency of the PA laser is swept while illuminating the ultracold mixture. In general, as we address deeper lying vibrational levels of the electronically excited states, the sweep time is extended to overcome the reduction in Franck-Condon factors. Typical sweep rates range between 1 GHz/sec to 10 GHz/sec. The center positions of the sweeps are recorded using a commercial wavemeter with a spectral resolution of 10 MHz and an absolute accuracy of about 300 MHz.

Photoassociation leads to a simultaneous loss of $^{23}$Na and $^{40}$K atoms from the dipole trap. After exposure, photoassociation resonances are located by counting the numbers of remaining Na and K atoms on absorption images of both species. The cycle time for a single set of absorption images is about 30 seconds.

\begin{figure}
  \begin{center}
  \includegraphics[width=1\columnwidth]{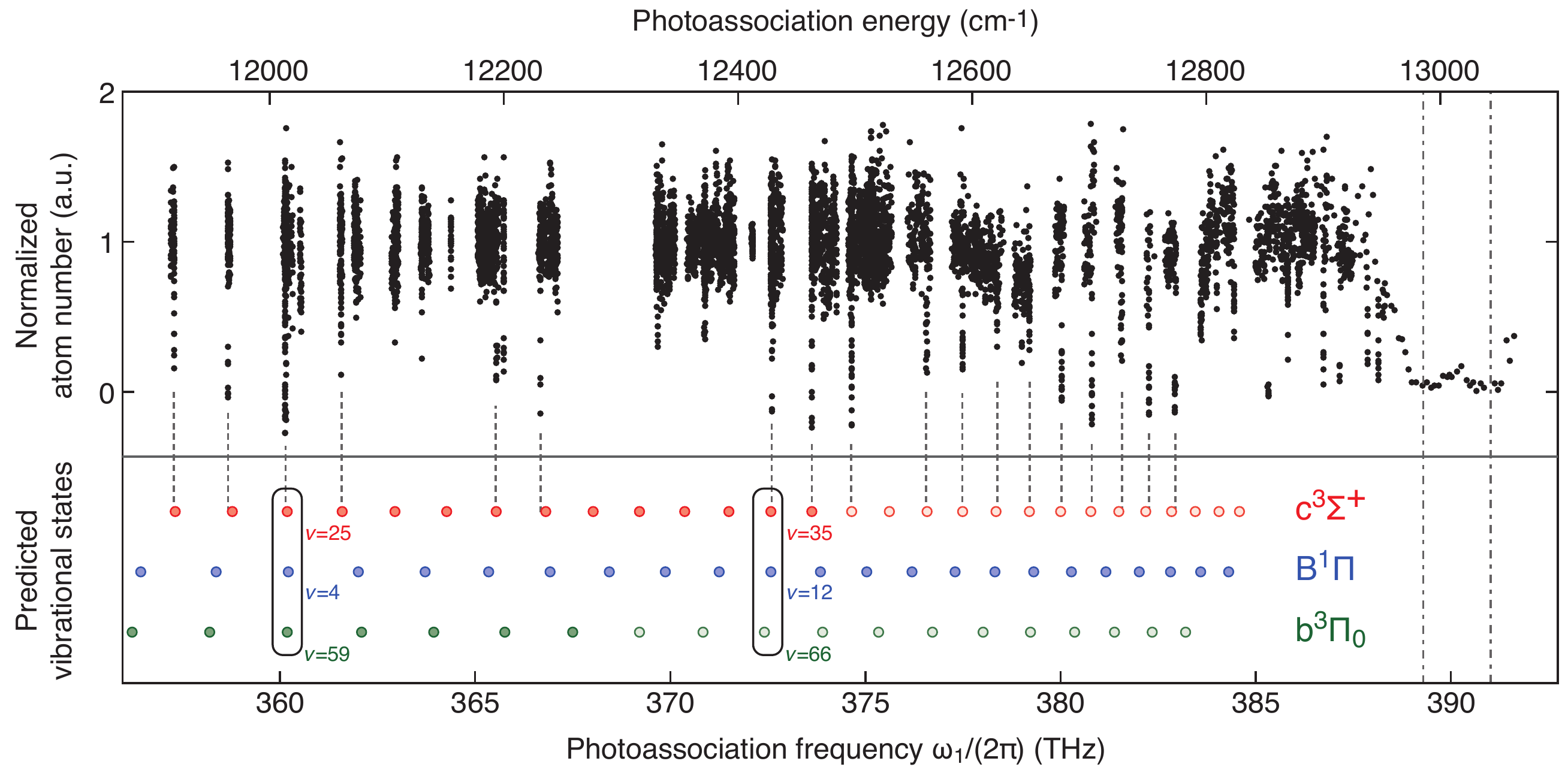}
  \end{center}
  \caption{\label{fig:PA} Photoassociation spectra accessing excited molecular states of $^{23}$Na$^{40}$K. The observed lines (upper panel) are compared to the prediction of vibrational levels (lower panel) of ${\rm B}^1 \Pi$ (blue), ${\rm c}^3 \Sigma^+$ (red), ${\rm b}^3\Pi_0$ (green), with lighter colors denoting predictions that lie outside the validity of known Dunham coefficients. Dashed vertical lines indicate the position of resonances that are assigned to the ${\rm c}^3\Sigma^+$ potential. The dot-dashed lines on the right show the $^{40}$K $D_1$ and $D_2$ transitions at 12985.2~cm$^{-1}$ and 13042.9~cm$^{-1}$, respectively.}
\end{figure}


We investigated electronically excited molecular states in the energy range from 11900~cm$^{-1}$ up to 13000~cm$^{-1}$, below the atomic $D_1$ transition of $^{40}$K. Figure \ref{fig:PA} shows a map of the recorded photoassociation spectra. In order to assign electronic states and vibrational quantum numbers to observed PA transitions, the measured transition frequencies are compared to predictions based on Dunham coefficients for $^{23}$Na$^{39}$K~\cite{Kasahara1991B1Pi,Ferber2000NaK,Ross1986b3Pi} that we mass-scaled for \nak.

We remind here that previous information on the location of triplet energy levels was available only through deperturbation of strongly perturbed levels mixing with singlet states. Errors beyond 1 cm$^{-1}$ are typical, as we see a posteriori from our study.
Using an ultracold mixture of unbound atoms, we instead directly access triplet states. Pairs of $^{23}$Na and $^{40}$K atoms in the $|1, 1 \rangle$ and $| 9/2, -9/2 \rangle$ states combine into a predominantly ``stretched" total spin state $|F,m_{F}\rangle = |7/2,-7/2\rangle$. Unbound atom pairs of $^{23}$Na and $^{40}$K thus have a dominant (\mbox{87.5~\%}) triplet character. Accordingly, most of the observed photoassociation resonances can be assigned to the ${\rm c}^{3}\Sigma^{+}$ potential. This becomes obvious by comparison to predicted positions of vibrational states of the relevant excited state potentials ${\rm b}^{3}\Pi, {\rm c}^3\Sigma^+$, and ${\rm B}^1\Pi$ (see Fig.~\ref{fig:PA}). The ${\rm b}^3\Pi$ state is expected to be less visible, as the transition dipole moment between ${\rm a}^3\Sigma^+$ and ${\rm b}^3\Pi$ vanishes at short-range~\cite{Aymar2007NaK}.
From the measured ${\rm c}^3\Sigma^+$ energy levels, it will be possible in future work to generate a highly accurate potential energy curve for this electronic state.

The photoassociation spectra allow us to identify two candidate states near $360$~THz and $372.5$~THz, in which vibrational levels of the ${\rm c}^3\Sigma^+$ and ${\rm B}^1\Pi$ potentials are close enough to display significant singlet-triplet mixing (see Fig.~\ref{fig:PA}).
The pairs in question are {${\rm B}^1\Pi |v{=}4 \rangle {\sim} {\rm c}^3\Sigma^+ | v{=}25 \rangle$} and {${\rm B}^1\Pi | v{=}12 \rangle {\sim}{\rm c}^3\Sigma^+| v{=}35 \rangle$}. In both cases, accidentally a third electronic state, ${\rm b}^3\Pi_\Omega$, intervenes, where the sub-states $\Omega=0,1,2$ are split by 15~cm$^{-1}$~\cite{Ross1986b3Pi} due to spin-orbit interaction. Here, $\Omega$ denotes the total orbital and spin angular momentum projection along the internuclear axis.

\begin{figure}
  \begin{center}
  \includegraphics[width=1\columnwidth]{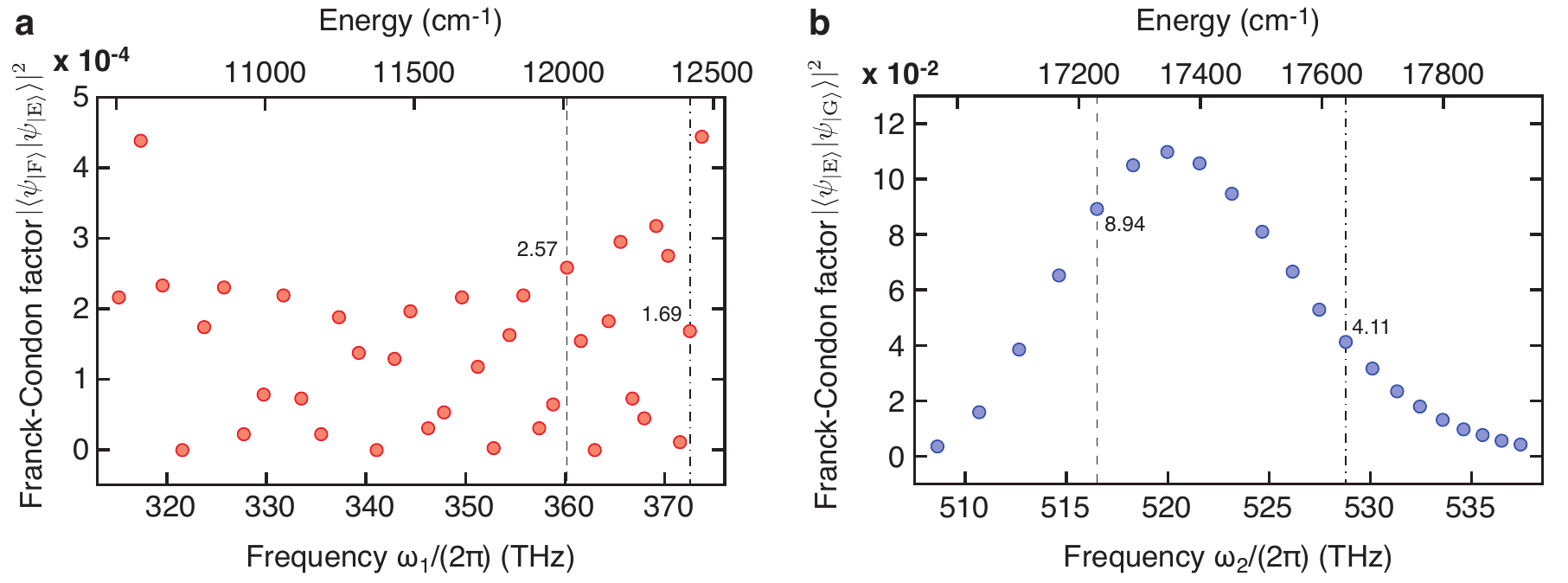}
  \end{center}
  \caption{\label{fig:FC} Franck-Condon factors between (a) the Feshbach level ${\rm a}^{3}\Sigma^{+}|v{=}19\rangle$ and excited vibrational levels of the ${\rm c}^3\Sigma^{+}$ potential, denoted by the transition frequency $\omega_1$, and (b)   excited vibrational levels of the ${\rm B}^1\Pi$ potential and the absolute ground state ${\rm X}^{1}\Sigma^{+}|v{=}0\rangle$, denoted by the transition frequency $\omega_2$. The Franck-Condon factor is defined as the squared absolute value of the wavefunction overlap. The vibrational wavefunctions are calculated for $^{23}$Na$^{40}$K using the molecular potentials known for $^{23}$Na$^{39}$K. The dashed and dot-dashed lines correspond to the ${\rm a}^{3}\Sigma^{+} | v {=} 25 \rangle {\sim} {\rm B}^{1}\Pi | v {=} 4 \rangle$ and the ${\rm c}^{3}\Sigma^{+} | v {=} 35 \rangle {\sim} {\rm B}^{1}\Pi | v {=} 12 \rangle$ vibrational levels, respectively (see text).
}
\end{figure}

For strong two-photon coupling, we require an intermediate state that strongly connects the initial Feshbach molecular state, dominantly associated with the triplet state ${\rm a}^{3}\Sigma^{+}$$|v{=}19, N{=}0, J{=}1\rangle$, to the absolute rovibrational singlet ground state ${\rm X}^{1}\Sigma^{+}$$|v{=}0, J{=}0\rangle$. Here, $N$ denotes the quantum number associated with the rotational angular momentum, and $J$ the total angular momentum neglecting nuclear spins. There are two criteria that need to be fulfilled: First, the intermediate state must feature strong singlet-triplet mixing to mediate coupling between the dominantly triplet Feshbach molecular state and the singlet rovibrational ground state. Second, the intermediate state should have a large transition dipole moment for coupling with both the Feshbach and the absolute ground state. Since the electronic part of the transition dipole moment is fairly constant for both the ${\rm a}^3\Sigma^+ \,{\rightarrow}\, {\rm c}^3\Sigma^+$ (${\approx} 10\,\rm D$) and the ${\rm B}^1\Pi \,{\rightarrow} \, {\rm X}^1\Sigma^+$ (${\approx} 7\,\rm D$) transitions~\cite{Aymar2007NaK}, the relevant figure of merit for the second criterion is the Franck-Condon (FC) overlap between the states. Fig.~\ref{fig:FC} shows the calculated Franck-Condon factors for the transitions from ${\rm a}^{3}\Sigma^{+}$$|v{=}19\rangle$ to ${\rm c}^{3}\Sigma^{+}$ and ${\rm B}^{1}\Pi$ to \mbox{${\rm X}^{1}\Sigma^{+}$$|v{=}0\rangle$}. The FC factors for both candidate states are significant. Note that the typical FC factor for the up-leg transition is two orders of magnitude smaller than for the down-leg transition, indicating that it will be more difficult to achieve substantial coupling between the Feshbach state and the excited state than between the excited state and the absolute ground state.

Regarding the criterion of strong singlet-triplet mixing, spin-orbit matrix elements $\xi_{\rm Bc}$ have been experimentally obtained for \Bcres with $^{23}$Na$^{39}$K~\cite{Kowal1989NaKtriplet,Ferber2000NaK}.
These works consistently found for {${\rm B}^1\Pi |v{=}4 \rangle {\sim} {\rm c}^3\Sigma^+ | v{=}25\rangle$} a value of $\xi_{\rm Bc}=0.16 \,\rm cm^{-1}$ and for {${\rm B}^1\Pi | v{=}12 \rangle {\sim}{\rm c}^3\Sigma^+ | v{=}35 \rangle$} $\xi_{\rm Bc}=0.58\,\rm cm^{-1}$. The spin-orbit matrix elements are approximately given by the electronic part of the spin-orbit coupling, which is fairly constant as a function of internuclear distance~\cite{Ferber2000NaK}, times the Franck-Condon overlap between the perturbing states. We checked that this overlap does not change significantly between $^{23}$Na$^{39}$K and $^{23}$Na$^{40}$K. Therefore, these values of $\xi_{\rm Bc}$ give an indication how close in energy the singlet and triplet states have to be to undergo significant mixing. In the next section, we investigate the strength of spin-orbit coupling for the candidate states above.


\section{Resonantly enhanced singlet-triplet mixing between ${\rm B}^1\Pi$ and ${\rm c}^3\Sigma^+$}
\label{s:mixing}
\begin{figure}
  \begin{center}
  \includegraphics[width=0.9\columnwidth]{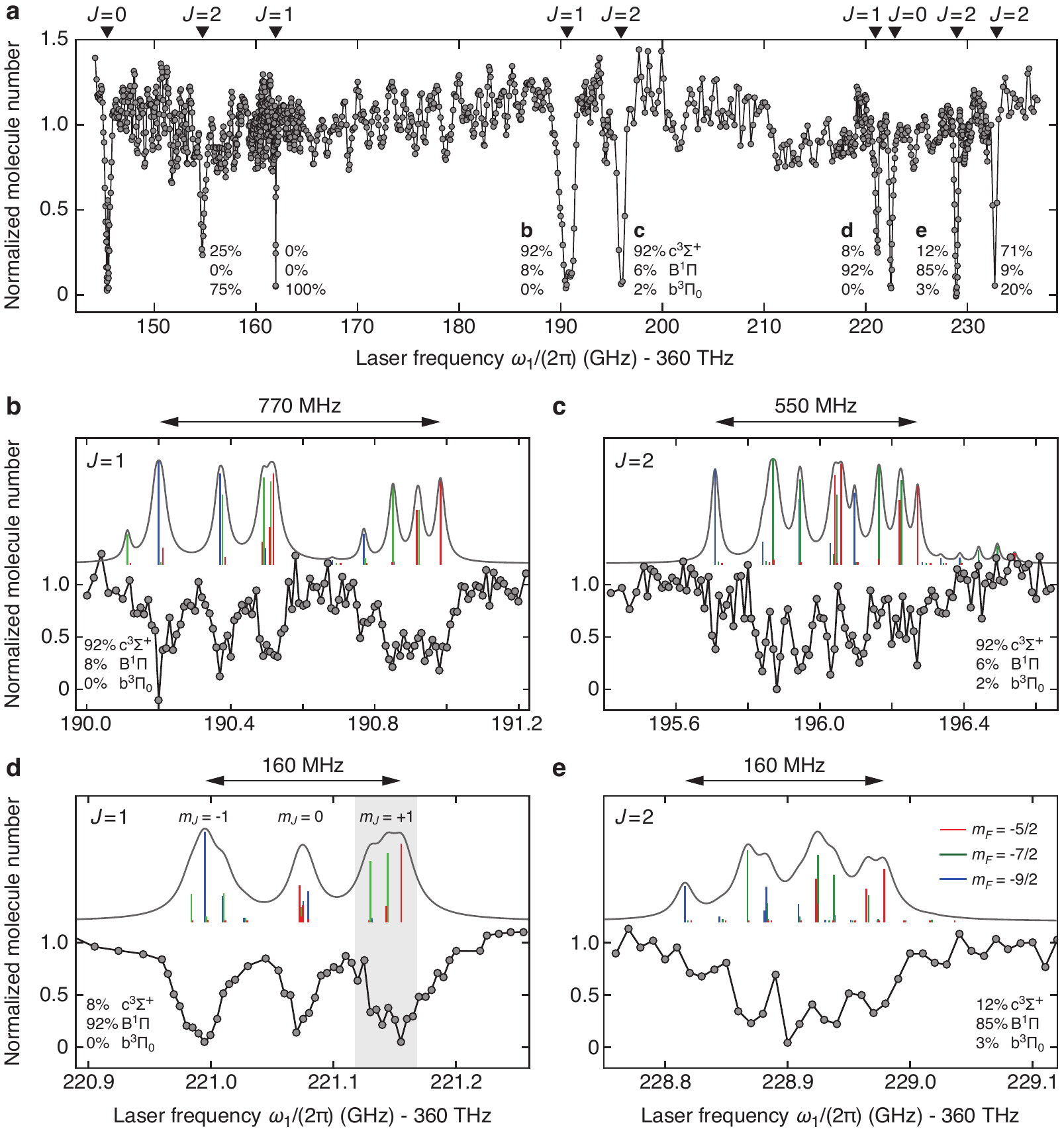}
  \end{center}
  \caption{\label{fig:finespectra832} High resolution single-photon spectroscopy on Feshbach molecules near 360~THz, exploring the coupled ${\rm B}^1\Pi\,|v{=}4\rangle {\sim} {\rm c}^3\Sigma^+\, |v{=}25\rangle {\sim} {\rm b}^3\Pi_0\,|v{=}59\rangle$ state manifold. (a) 90 GHz wide scan exposing a rich fine structure of nine loss features. The indicated loss features at $360.190\,\rm THz$ and $360.196\,\rm THz$ are dominantly ${\rm c}^3\Sigma^+$ states with $J{=}1$ and $J{=}2$, respectively. Their hyperfine structure is shown in high resolutsion scans of panels (b) and (c). The two features at $360.221\,\rm THz$ and $360.229\,\rm THz$ correspond to the dominantly singlet ${\rm B}^1\Pi |J{=}1\rangle$ and ${\rm B}^1\Pi  | J{=}2\rangle$ level, respectively. Their Zeeman and hyperfine structure is shown in panels (d) and (e). In (b)-(e), the gray solid lines represent a calculation of the hyperfine structure, including the Zeeman interaction (see text). The blue, green and red lines indicate the spectroscopic weight of the individual contributing hyperfine features with $m_F {= -}9/2$, $-7/2$ and $-5/2$, respectively, corresponding to $\sigma^-$, $\pi$ and $\sigma^+$ light. The gray shading in (d) marks the $m_J {= +} 1$ loss feature that is used for ground state spectroscopy (see Figs.~\ref{fig:twophoton} and \ref{fig:EIT}). The relevant sub-features (green) have quantum numbers $m_{F_1}{=}-1/2$ (left) and $m_{F_1}{=}1/2$ (right).}
\end{figure}

\begin{figure}
  \begin{center}
  \includegraphics[width=0.9\columnwidth]{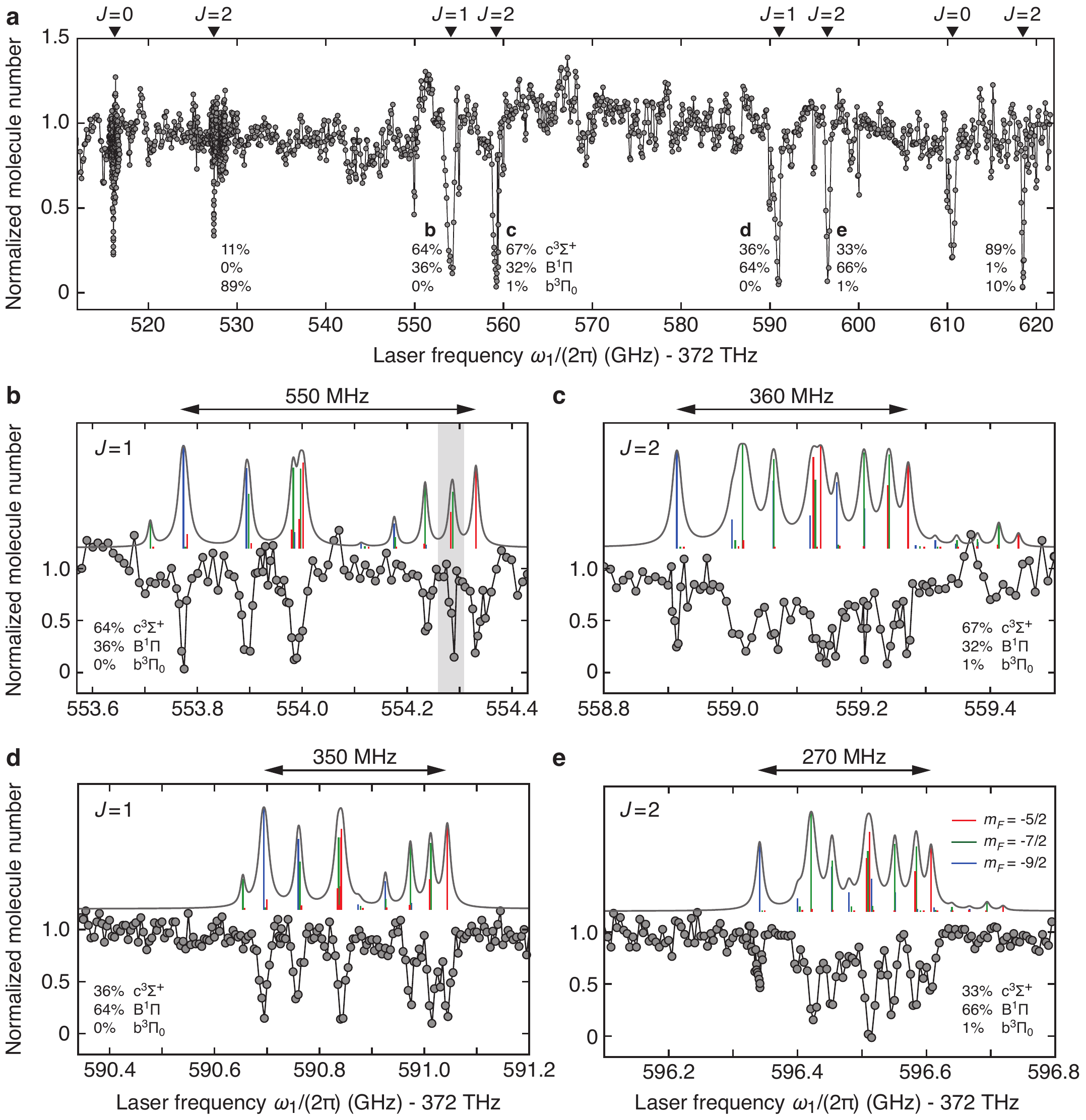}
  \end{center}
  \caption{\label{fig:finespectra804}
  High resolution single-photon spectroscopy near 372.5 THz, exploring the coupled ${\rm B}^1\Pi\,|v{=}12\rangle {\sim} {\rm c}^3\Sigma^+\, |v{=}35\rangle{\sim} {\rm b}^3\Pi_0\,|v{=}66\rangle$ state manifold. (a) A 100 GHz scan reveals eight loss features. (b) and (d) displays the hyperfine structure of the two $J{=}1$ states. Significant singlet-triplet mixing is evident due to the similarity of the observed features. The differing width of the hyperfine structure indicates the differing admixture of the hyperfine-rich ${\rm c}^3\Sigma^+$ state. Likewise, the $J{=}2 $ states shown in (c) and (e) share the same hyperfine features. Gray solid lines, and blue, green and red lines in (b)-(e) as in Fig.~\ref{fig:finespectra832}. The gray shading in (b) marks the sublevel that is used for dark state spectroscopy (see Fig.~\ref{fig:EIT}). The relevant sub-feature (green) has quantum numbers $F_1{=}5/2$ and $m_{F_1}{=}1/2$. }
\end{figure}

\subsection{Experimental procedure}

In order to further investigate the rotational, fine, and hyperfine structure of the two candidate state manifolds, we perform high resolution one-photon spectroscopy on $^{23}$Na$^{40}$K Feshbach molecules. Working with Feshbach molecules in contrast to free atoms enhances the coupling to the excited states, and deeply bound vibrational levels can be efficiently addressed. Instead of seconds, resonant excitation of Feshbach molecules typically occurs on a time scale of tens of microseconds.

To create Feshbach molecules we use radiofrequency (rf) association as described in~\cite{wu2012NaK}. In short, after preparation of the mixture in the optical dipole trap, $^{40}$K atoms are transferred to the $|9/2, -7/2\rangle$ state. Then, a uniform magnetic field of 85.7 G is applied to the atoms, close to a $s$-wave Feshbach resonance in the $|f_{\rm Na}{=}1,m_{f_{\rm Na}}{=}1,f_{\rm K}{=}9/2,m_{f_{\rm K}}{=-}9/2\rangle$ scattering channel caused by the closed channel molecular state ${\rm a}^3\Sigma^+ |v{=}19,N{=}0,J{=}1,F{=}9/2,m_F{=-}7/2\rangle$. Here, $F$ is the total molecular angular momentum including the Na and K nuclear spins, and $m_F$ its projection along the quantization axis. The quantization axis is set by the magnetic field in the vertical direction ($z$-axis).
We associate Feshbach molecules by driving the free-bound transition from  $|f_{\rm Na}{=}1,m_{f_{\rm Na}}{=}1,f_{\rm K}{=}9/2,m_{f_{\rm K}}{=-}7/2\rangle$ to the molecular state causing the Feshbach resonance in $|f_{\rm Na}=1,m_{f_{\rm Na}}=1,f_{\rm K}=9/2,m_{f_{\rm K}}{=-}9/2\rangle$, with the additional binding energy being provided by the rf photon.
At the given magnetic field, we measure the binding energy of the Feshbach molecules to be 80 kHz and convert about 10\% of the atoms to yield about $10^4$ Feshbach molecules.

The associated Feshbach molecules are illuminated by laser 1 (see Fig.~\ref{fig:cartoon}) to drive transitions to excited states of $^{23}$Na$^{40}$K. Resonant coupling leads to a loss of Feshbach molecules. The intensity of the laser and the exposure time are chosen to clearly reveal the sub-structure of the excited states without saturating the transitions. The laser polarization is fixed to be diagonal with respect to the bias magnetic field. After exposure to the probe laser, an absorption image is taken and the number of remaining Feshbach molecules counted as a function of laser detuning. For absorption imaging, we use light resonant with the $^{40}$K atomic transition. Absorption of a first photon breaks the weakly bound Feshbach molecules, followed by scattering of additional photons on the atomic transition.

To achieve high spectral resolution required for precision spectroscopy, the frequency of the probe laser is locked to a high-finesse ultralow expansion (ULE) cavity in a master-slave configuration. The master laser is directly locked to a transmission mode of the ULE cavity by a Pound-Drever-Hall lock. The probe laser is phase-locked to the master laser with a variable frequency offset. For both lasers, we use home-built grating stabilized diode lasers. This configuration allows flexible adjustment of the probe laser detuning over a range of 10~GHz with kHz precision. The cavity has a finesse of $\sim$35,000 near 360.2~THz (832.2~nm), and $\sim$15,000 near 372.6~THz (804.7~nm), suitable for addressing both candidate states. The temperature of the cavity is stabilized for long term frequency stability. Absolute frequency determination is performed for the most part with the help of a commercial wavemeter of 300~MHz accuracy as described above. One particular single-photon transition is determined to 3~MHz accuracy using a frequency comb, as discussed below.

\subsection{Analysis of the fine structure}

\begin{figure}
  \begin{center}
  \includegraphics[width=.6\columnwidth]{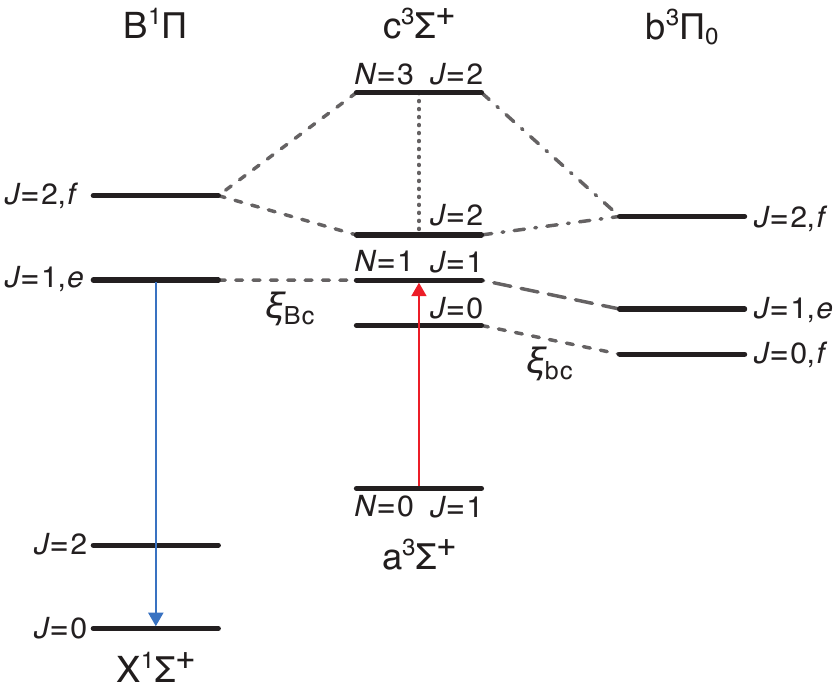}
  \end{center}
  \caption{\label{fig:statediagram}
Diagram of deperturbed states relevant for the chosen two-photon pathway connecting the Feshbach state in ${\rm a}^3\Sigma^+$ to the singlet ground state ${\rm X}^1\Sigma^+$. For the ground (electronically excited) state manifolds, only positive (negative) parity states are relevant and shown here. Accordingly, electronically excited states with odd (even) $J$ have $e$ ($f$) parity. ${\rm B}^1\Pi$ states are coupled via spin-orbit coupling (dashed lines) to ${\rm c}^3\Sigma^+$, which in turn is coupled to ${\rm b}^3\Pi_0$ via spin-orbit coupling (short dashes), L-uncoupling (long dashes) or both (dot-dashed line). The vertical dotted line connecting ${\rm c}^3\Sigma^+$ $|N{=}3,J{=}2\rangle$ and $|N{=}1,J{=}2\rangle$ represents spin-spin coupling. Transitions used for two-photon coupling to the $J{=}0$ absolute rovibrational ground state are indicated. To reach $J{=}2$ of the singlet ground state, one can either choose $J{=}1$ or $J{=}2$ in the intermediate excited state.}
\end{figure}

Figures \ref{fig:finespectra832} and~\ref{fig:finespectra804} show the observed fine (a) and hyperfine (b)-(e) structure of the two state manifolds at 360~THz and 372~THz, respectively. In order to understand the spectral structure, we analyze the perturbations among nearly degenerate electronically excited states, following the approach of Ref.~\cite{Ishikawa1992c3shyperfine}. As a result of the theoretical analysis presented here, we are able to reproduce the observed fine and hyperfine structure, to characterize the electronic state decomposition, and to assign angular momentum quantum numbers.

The fine structures of the 360~THz and 372~THz manifolds in Figs.~\ref{fig:finespectra832}(a) and \ref{fig:finespectra804}(a) show nine and eight significant loss features, respectively. The state diagram of Fig.~\ref{fig:statediagram} illustrates the various molecular states that are involved in the fine structure. Starting with non-rotating triplet Feshbach molecules in a $N{=}0, J{=}1$ state of positive parity, only electronically excited states with negative parity and $J{=}0$, $J{=}1$ or $J{=}2$ can be reached. The ${\rm c}^3\Sigma^+$ electronic state, being close to Hund's case b), offers four such states: $|N{=}1,J{=}(0,1,2)\rangle$ and $|N{=}3,J{=}2\rangle$, where the latter is coupled to $|N{=}1,J{=}2\rangle$ via spin-spin coupling (vertical dotted line in Fig.~\ref{fig:statediagram}). The ${\rm B}^1\Pi$ electronic state contributes two states, the $e$ state with $J{=}1$ and the $f$ state with $J{=}2$. The remaining three (360 THz manifold) and two (372 THz manifold) features are associated with the ${\rm b}^3\Pi_0$ electronic state, introducing three states with $J{=}0$, $J{=}1$, and $J{=}2$. As only negative parity states are involved in the excited state, odd $J$ corresponds to $e$ parity and even $J$ to $f$ parity. The $e$/$f$ label will be omitted in the following discussion. 

The fine structure around $360\,\rm THz$ is associated with the perturbed complex
${{\rm B}^1\Pi |v{=}4 \rangle{\sim} {\rm c}^3\Sigma^+ | v{=}25 \rangle {\sim} {\rm b}^3\Pi_0 |v{=}59\rangle} $, while the $372\,\rm THz$ manifold is caused by 
${\rm B}^1\Pi |v{=}12 \rangle {\sim}{\rm c}^3\Sigma^+ | v{=}35 \rangle {\sim}{\rm b}^3\Pi_0 |v{=}66\rangle $. For clarity of this presentation, we neglect the influence of the other ${\rm b}^3\Pi_{1,2}$ states that are far off-resonant, a distance of $15\,\rm cm^{-1}$ and $30\,\rm cm^{-1}$ higher in energy \cite{Ross1986b3Pi}. However, they are included in the calculations that reproduce the observed spectra, where they cause minor shifts of lines. 

The Hamiltonian governing the fine structure of the perturbed ${\rm B}{\sim}{\rm c}{\sim}{\rm b}$ complex is~\cite{Ishikawa1992c3shyperfine,Field2004spectra,Brown:2003}
\begin{equation}
  H_{\rm fine} = H^{\rm B}_0+H^{\rm c}_0+H^{\rm c}_{\rm SS}+H^{\rm c}_{\rm SR}+H^{\rm b}_0+H^{\rm Bc}_{\rm SO}+H^{\rm bc}_{\rm SO}+H^{\rm bc}_{\rm BL}.
\end{equation}
The deperturbed energy levels $E_{\rm B, c, b}$ of electronic states and their rotational energies (with rotational constants $B_{\rm B,c,b}$) are included in $H^{\rm B, c, b}_0$. $E_{\rm B}$ and $E_{\rm c}$ are taken to be the deperturbed energies in the absence of spin-orbit coupling of ${\rm B}^1\Pi |J{=}1\rangle$ and ${\rm c}^3\Sigma^+ |N{=}1,J{=}1\rangle$, respectively. $E_{\rm b}$ is the deperturbed energy of the ${\rm b}^3\Pi_0\, | J{=}0\rangle$ state. Within ${\rm c}^3\Sigma^+$, we include spin-spin coupling $H^c_{\rm SS}$ with coupling constant $\lambda$, and the weak spin-rotation coupling $H^c_{\rm SR}$ with constant $\gamma$, which is diagonal in the chosen Hund's case b) basis. The couplings are schematically shown in Fig.~\ref{fig:statediagram}. States ${\rm B}^1\Pi$ and ${\rm c}^3\Sigma^+$ are connected via spin-orbit coupling $H^{\rm Bc}_{\rm SO}$ (coupling constant $\xi_{\rm Bc}$). Also the electronic states ${\rm b}^3\Pi_0$ and ${\rm c}^3\Sigma^+$ are coupled via spin-orbit coupling $H^{\rm bc}_{\rm SO}$ (coupling constant $\xi_{\rm bc}$), and in addition, for $J\ne 0$, via weak $L$-uncoupling $H^{\rm bc}_{\rm BL}$ (coupling constant $B_L$). The later has a minute influence for the low rotational states considered here, and will be neglected in the discussion, but is included in the calculation.

All of the above perturbations conserve total angular momentum $J$, so that the full Hamiltonian can be broken down into sub-Hamiltonians that only act within sub-spaces with fixed $J$.
In the basis $\{ {\rm c}^3\Sigma^+ |N{=}1,J{=}0\rangle,{\rm b}^3\Pi_0 |J{=}0\rangle \}$, the Hamiltonian for $J{=}0$ is
\begin{equation*}
H_{J=0}=\left(
  \begin{array}{cc}
    E_{\rm c}-\gamma-2\lambda & -\sqrt{2}\xi_{\rm bc} \\
    -\sqrt{2} \xi_{\rm bc} & E_{\rm b} \\
  \end{array}
\right).
\end{equation*}
For $J{=}1$, neglecting $L$-uncoupling, another $2{\times}2$ matrix connects the states in the basis
$\{ {\rm B}^1\Pi |J{=}1\rangle, {\rm c}^3\Sigma^+ |N{=}1,J{=}1\rangle \}$ via spin-orbit coupling,
\begin{equation*}
H_{J=1}=\left(
  \begin{array}{cc}
    E_{\rm B}& \xi_{\rm Bc} \\
    \xi_{\rm Bc} & E_{\rm c} \\
  \end{array}
\right).
\end{equation*}
$L$-uncoupling causes weak coupling of $ {\rm c}^3\Sigma^+ |N{=}1,J{=}1\rangle$ to ${\rm b}^3\Pi_0 |J{=}1\rangle$, which is responsible for the extremely weak and narrow $J{=}1$ feature near 360.162~THz in Fig.~\ref{fig:finespectra832}(a). The corresponding feature in the 372 THz manifold was not found, presumably due to even weaker $L$-uncoupling.
Finally, the $J{=}2$ sub-space is governed by a $4{\times}4$ matrix in the basis $\{ {\rm B}^1\Pi |J{=}2\rangle, {\rm c}^3\Sigma^+ |N{=}1,J{=}2\rangle, {\rm c}^3\Sigma^+ |N{=}3,J{=}2\rangle, {\rm b}^3\Pi_0 |J{=}2\rangle \}$, neglecting influence of far off-resonant states ${\rm b}^3\Pi_{1,2}$,
\begin{equation*}
H_{J=2}=
\left(
  \begin{array}{cccc}
    E_{\rm B}+4 B_{\rm B} & \sqrt{\frac{3}{5}} \xi_{\rm Bc} & \sqrt{\frac{2}{5}} \xi_{\rm Bc} & 0 \\
    \sqrt{\frac{3}{5}} \xi_{\rm Bc} & E_{\rm c}+2\gamma-\frac{4}{5}\lambda & \frac{2\sqrt{6}}{5}\lambda & \sqrt{\frac{4}{5}}\xi_{\rm bc} \\
    \sqrt{\frac{2}{5}} \xi_{\rm Bc}  & \frac{2\sqrt{6}}{5}\lambda & E_{\rm c}+10 B_{\rm c}-3 \gamma-\frac{6}{5}\lambda & -\sqrt{\frac{6}{5}} \xi_{\rm bc} \\
    0 & \sqrt{\frac{4}{5}}\xi_{\rm bc} & -\sqrt{\frac{6}{5}} \xi_{\rm bc} & E_{\rm b}+6 B_{\rm b} \\
  \end{array}
\right).
\end{equation*}

As a starting point for the analysis of the spectra, we use deperturbed energy levels and rotational constants, obtained by mass-scaling of existing $^{23}$Na$^{39}$K data on ${\rm B}^1\Pi$~\cite{Kato1990c3s}, ${\rm c}^3\Sigma^+$~\cite{Kowal1989NaKtriplet,Ferber2000NaK} and ${\rm b}^3\Pi_0$~\cite{Ross1986b3Pi,Kowal1989NaKtriplet,Ferber2000NaK} states. ${\rm b}^3\Pi_0 |v=66\rangle$ has not been observed before and falls outside the validity of the Dunham expansion given in~\cite{Ferber2000NaK}. The above perturbations shift the energy levels, and the corresponding coupling constants serve as variables to match the theory with the experimental data. An initial estimate of $\xi_{\rm Bc}=0.58\,\rm cm^{-1}$ for the 360 THz manifold and $0.16\,\rm cm^{-1}$ for the 372 THz manifold was discussed in the previous section~\cite{Ferber2000NaK}. The value of $\xi_{\rm bc} = 0.77$ for the 360 THz manifold is obtained from measured values in~\cite{Ferber2000NaK} after weighting with the proper Franck-Condon factors. Values of $\gamma$ and $B_L$ for $^{23}$Na$^{39}$K can be found in~\cite{Kowal1989NaKtriplet} and serve as initial values here. By varying the spin-spin coupling constant $\lambda$ of ${\rm c}^3\Sigma^+$, the unknown spin-orbit coupling $\xi_{\rm bc}$ of the 372 THz manifold, and the deperturbed energy levels $E_{\rm B, c, b}$, excellent agreement with the observed fine and hyperfine structure is obtained (see Figs.~\ref{fig:finespectra832} and~\ref{fig:finespectra804}).

The $J$ quantum numbers of the observed fine structure features are identified by considering their distinct hyperfine and Zeeman sub-structure. The two $J{=}0$ states are featureless, and $J{=}1$ and $J{=}2$ states show characteristic patterns, to be analyzed below. For either of the two manifolds, strong ${\rm b}{\sim}{\rm c}$ spin-orbit coupling explains the unusual presence of a $J{=}0$ state in the midst of states with higher $J$. 

For the present task of finding a two-photon pathway from Feshbach molecules to the absolute singlet ground state, the $J{=}1$ states are highly relevant. In the 360~THz manifold (see Fig.~\ref{fig:finespectra832}(a)) the two $J{=}1$ features are well separated by 1~cm$^{-1}$. This is significantly larger than the expected strength of spin-orbit coupling for these states given the initial estimate of $\xi_{\rm Bc}=0.16\,\rm cm^{-1}$ (our analysis yields $0.27\,\rm cm^{-1}$, see below). Thus the states are relatively weakly mixed. As a consequence, the high resolution scans of the two $J{=}1$ features in Fig.~\ref{fig:finespectra832}(b) and (d) show weak resemblance in their structure. Nevertheless, the very fact that two $J{=}1$ features are observable is a consequence of spin-orbit coupling, because a purely singlet state would not be detectable in our scans. The analysis reveals that state (b) has 92\% triplet character, and therefore displays a characteristic triplet hyperfine structure; state (d) has 92\% ${\rm B}^1\Pi$ character, and accordingly the most prominent feature is a Zeeman triplet of energy levels with magnetic quantum numbers $m_J {= -}1$, 0, and $+1$.

In stark contrast, the 372 THz manifold (see Fig.~\ref{fig:finespectra804}(a)) contains two pairs of states, two $J{=}1$ states ((b) and (d)) and two $J{=}2$ states ((c) and (e)), whose respective hyperfine structure is nearly identical. This ``line doubling'' of an apparent ${\rm c}^3\Sigma^{+}$ state is well-known from spectroscopic studies of NaK~\cite{Breford1981NaKperturbation,Barrow1987B1Pi,Baba1988NaK,Kowal1989NaKtriplet,Kato1990c3s,Ishikawa1992c3shyperfine}, and it results from strong mixing of ${\rm c}^3\Sigma^{+}$ with ${\rm B}^1\Pi$. The energy difference between the two $J{=}1$ features is about $1.23\,\rm cm^{-1}$, comparable to twice the spin-orbit matrix element $\xi_{\rm Bc} = 0.58$.
As the $|J{=}1\rangle$ states can approximately be considered to form a two-level system (see $H_{J=1}$ above), we can write each of them as a superposition:
\begin{eqnarray}
 \left|J{=}1\right>_{(b)} = \alpha \left|{\rm c}^3\Sigma^{+},N{=}1,J{=}1\right> + \beta \left|{\rm B}^1\Pi,J{=}1\right>\\
 \left|J{=}1\right>_{(d)} = -\beta^* \left|{\rm c}^3\Sigma^{+},N{=}1,J{=}1\right> + \alpha^* \left|{\rm B}^1\Pi,J{=}1\right>
 \end{eqnarray}
As the hyperfine structure is largely dominated by the ${\rm c}^3\Sigma^{+}$ contribution, the ratio of the widths of the hyperfine features reveals the value of $\left|\alpha\right|^2/\left|\beta\right|^2 \approx 1.6$. The full perturbation analysis including hyperfine and Zeeman structure gives $\left|\alpha\right|^2/\left|\beta\right|^2 = 1.8$, implying $\alpha =0.80$ and $\beta = 0.60$. We thus have identified a pair of states with a nearly even, 64\%-36\% mixing between singlet and triplet states.

\subsection{Analysis of the hyperfine structure}

To understand the hyperfine and Zeeman structure found in the high-resolution scans of Figs.~\ref{fig:finespectra832}(b)-(e) and Figs.~\ref{fig:finespectra804}(b)-(e), we include in the total Hamiltonian the Zeeman interaction of states ${\rm B}^1\Pi$ and ${\rm c}^3\Sigma^+$ ($H^{\rm B,c}_{\rm Z}$) and the hyperfine interaction within ${\rm c}^3\Sigma^+$, consisting of the Fermi contact interaction of the electronic spin with the sodium and the potassium nucleus ($H^{\rm c}_{\rm hf, Na}$ and $H^{\rm c}_{\rm hf, K}$) as described in Ref.~\cite{Ishikawa1992c3shyperfine},
\begin{equation*}
H_{\rm total} = H_{\rm fine} + H^{\rm B}_{\rm Z} +H^{\rm c}_{\rm Z}+H^{\rm c}_{\rm hf, Na}+H^{\rm c}_{\rm hf, K}.
\end{equation*}
The Zeeman Hamiltonian for the ${\rm B}^1\Pi$ state is~\cite{Brown:2003}
\begin{equation*}
  H^{\rm B}_{\rm Z} = \mu_B L_z B, 
\end{equation*}
where $\mu_B$ is the Bohr magneton and $L_z$ the $z$-component of the molecule's orbital angular momentum in the laboratory frame.
$H^{\rm B}_{\rm Z}$ is diagonal in the $|J m_J\rangle$ basis of rotational states of $B^1\Pi |v\rangle$ with matrix elements
\begin{equation}
  \left<J m_J\right|H^{\rm B}_{\rm Z}\left|J m_J\right>=\mu_B B \frac{m_J}{J(J+1)},
\end{equation}
describing the projection of the one unit of angular momentum along the internuclear axis onto the magnetic field in the laboratory $z$-axis; $m_J$ denotes the $z$-projection of $\vec{J}$. In particular, the magnetic moment of a $|J{=}1,m_J{=}1\rangle$ state is $\mu_B/2$, causing a Zeeman splitting of  about $ 60\,\rm MHz$, close to what we observe for  the dominantly ${\rm B}^1\Pi$ state in Fig.~\ref{fig:finespectra832}(d). Additional splitting and broadening is due to the admixture of ${\rm c}^3\Sigma^+$, which introduces hyperfine and additional Zeeman shifts.
For the ${\rm c}^3\Sigma^+$ state, the Zeeman term is
\begin{equation*}
  H^{\rm c}_{\rm Z} = \mu_B g_S S_z B
\end{equation*}
 with $S_z$ the $z$-component of the electron spin the laboratory frame, and $g_S$ its $g$-factor. Note that $H^{\rm c}_{\rm Z}$ is not diagonal in the Hund's case b) basis, which mixes states of differing spin projection along the internuclear axis~\cite{Kato:1984,Brown:2003}.
The hyperfine Hamiltonians are
\begin{equation*}
  H^{\rm c}_{\rm hf, Na} = \alpha_{\rm Na} \vec{I_{\rm Na}}\cdot\vec{S} \quad \mathrm{and} \quad H^{\rm c}_{\rm hf, K} = \alpha_{\rm K} \vec{I_{\rm K}}\cdot\vec{S},
\end{equation*}
where $\vec{I_{\rm Na}}$ and $\vec{I_{\rm K}}$ denote the nuclear spin for $^{23}$Na and $^{40}$K, respectively. A major simplification of the analysis is afforded by the large splitting between states of different $J$, compared to the hyperfine (and Zeeman) splitting ($\alpha_{\rm Na} \approx 340\,\rm MHz$ and $\alpha_{\rm K} \approx 30 \,\rm MHz$). $J$ therefore remains a good quantum number.

\begin{table}
\centering
\begin{tabular}{c | c | c}
  Parameter &  ${{\rm B} |v{=}4 \rangle{\sim} {\rm c}| v{=}25 \rangle {\sim} {\rm b} |v{=}59\rangle} $ & ${\rm B} |v{=}12 \rangle {\sim}{\rm c} | v{=}35 \rangle {\sim}{\rm b} |v{=}66\rangle $ \\
  \hline\hline
  $E_{\rm B}$ & 17289.158 & 17701.427\\
  $E_{\rm c}$  & 17288.338 & 17701.074\\
  $E_{\rm b}$ & 17287.308 & 17699.680\\
  $B_{\rm B}$ & 0.0662 & 0.0551\\
  $B_{\rm c}$ & 0.0461 & 0.0392\\
  $B_{\rm b}$ & 0.0624 & 0.0540\\
  $\xi_{\rm Bc}$ & 0.27 & 0.5899 ($J{=}1$), 0.5833 ($J{=}2$)\\
  $\xi_{\rm bc}$ & 0.766 & 0.68\\
  $\lambda$ & -0.174 & -0.55\\
  $\gamma$ & 0.0022 & 0.0215\\
  $B_L$ & 0.0035 & 0.006\\
  $\alpha_{\rm Na}$& 0.0112 & 0.0114\\
  $\alpha_{\rm ^{40}K}$ & -1.2 $10^{-3}$ & -0.94 $10^{-3}$\\
  \hline
\end{tabular}
\caption{\label{table:results} Molecular constants and interaction parameters for the two resonant ${\rm B}^1\Pi{\sim}{\rm c}^3\Sigma^+{\sim}{\rm b}^3\Pi_0$ manifolds. Calculated line positions of the fine structure agree with the experimental data to within the experimental accuracy of ${\sim} 300\,\rm MHz$, dominated by wavemeter error. All values are in units of $\rm cm^{-1}$.}
\end{table}

The dominant contribution of the hyperfine interaction in ${\rm c}^3\Sigma^+$ originates from the $^{23}$Na nucleus, which has a nuclear spin $I_{\rm Na} {= }3/2$. This leads to three (four) hyperfine manifolds for $J{=}1$ ($J{=}2$) at zero field, labelled by the quantum number $F_1$, corresponding to the addition of the sodium nuclear spin to the total angular momentum $\vec{J}$ excluding nuclear spins, $\vec{F}_1 = \vec{J} + \vec{I}_{\rm Na}$. For $J{=}1$, possible values are $F_1 = 1/2$, $3/2$, and $5/2$; for $J{=}2$ additionally $F_1 = 7/2$ occurs. The appropriate hyperfine constants for various vibrational levels of ${\rm c}^3\Sigma^+$, including the ones relevant here ($|v{=}25\rangle$ and $|v{=}35\rangle$) were measured in Refs.~\cite{Kowal1989NaKtriplet,Kato1990c3s,Ishikawa1992c3shyperfine} for $^{23}$Na$^{39}$K. To include the hyperfine interaction of electron spins with the potassium nucleus, one introduces the total angular momentum $\vec{F}=\vec{F}_1 + \vec{I}_{\rm K} =\vec{J} + \vec{I}_{\rm Na}+ \vec{I}_{\rm K}$, with $I_{\rm K} = 4$, causing new levels to appear from each $F_1$ level (for example, $F_1{=}1/2$ leads to $F{=}7/2$ and $F{=}9/2$). Finally, the Zeeman interaction at $85.7 \,\rm G$ further splits these lines. The only remaining strictly ``good'' quantum number is the $z$-component of the total angular momentum, $m_{F}$. Still, for well-separated levels, $F$ or $F_1$ and the $z$-component $m_{F_1}$ are often approximately defined.

In order to understand the observed hyperfine spectra, we analyze which states are accessible starting from the initial Feshbach molecular state with $m_F{= -}7/2$. Possible final states that can be reached via diagonal polarization are $m_{F} {= -}5/2,-7/2,-9/2$. An additional approximate selection rule comes from the fact that the closed-channel Feshbach molecular state is predominantly in the total angular momentum state $F{=}9/2$, the open-channel is in $F{=}7/2$, so that transitions to excited states $F{=}13/2$ or $F{=}3/2$ are forbidden. Working in the nuclear spin-decoupled basis of $|N,J,m_J,I_{\rm Na},m_{I_{\rm Na}},I_{\rm K},m_{I_{\rm K}}\rangle$ for ${\rm c}^3\Sigma^+$, and noticing that the Feshbach molecules have $|N{=}0, J{=}1\rangle$, the line strength is described by the H\"onl-London factor $\frac{2J+1}{9}|\left<J m_J |1 p 1 m_J'\right>|^2$. The symbol in brackets denotes a Clebsch-Gordan matrix element, and $p$ describes the laser polarization, where $p=0$ stands for linear, and $p=\pm 1$ for $\sigma^\pm$ circularly polarized light. The calculated spectra are shown above the measured data in Fig.~\ref{fig:finespectra832} (b)-(e) and Fig.~\ref{fig:finespectra804} (b)-(e). They reproduce the observed line structure well. Most lines consist of several individual hyperfine components, indicated by individual lines within the theoretical spectrum. Table~\ref{table:results} summarizes the parameters used to reproduce the observed spectra.

We can compare our results for the hyperfine constants with the literature on $^{23}$Na$^{39}$K. For ${\rm c}^3\Sigma^+ |v{=}25\rangle$ and $|v{=}35\rangle$, Ref.~\cite{Kowal1989NaKtriplet} found $\alpha_{\rm Na} = 312(6)\,\rm MHz$ (in agreement with~\cite{Ishikawa1992c3shyperfine}) and $342(6)\,\rm MHz$, respectively. For $^{23}$Na$^{40}$K, our analysis yields $\alpha_{\rm Na} = 334(4) \,\rm MHz$ for $|v{=}25\rangle$ and $\alpha_{\rm Na} = 342(2) \,\rm MHz$ for $|v{=}35\rangle$, the latter in excellent agreement with the work on $^{23}$Na$^{39}$K~\cite{Kowal1989NaKtriplet}.
The influence of the potassium nucleus on the hyperfine structure had previously either not been observed~\cite{Kowal1989NaKtriplet} or only indirectly through broadening of Na-dominated hyperfine lines~\cite{Ishikawa1992c3shyperfine}, leading to estimates $\alpha_{\rm ^{39}K} < 10\,\rm MHz$ and $18\,\rm MHz$, respectively. From the known atomic hyperfine constant, the probability for finding an $s$-electron near the $^{39}$K nucleus was estimated to be 15\%~\cite{Ishikawa1992c3shyperfine}. For $^{40}$K, this would yield $\alpha_{\rm ^{40}K} = - 22\,\rm MHz$; note the negative sign due to the anomalous sign of the $g$-factor of the $^{40}$K nucleus. Here, we find $\alpha_{\rm ^{40}K}{=}-37(4)\,\rm MHz$ for $|v{=}25\rangle$ and $\alpha_{\rm ^{40}K}{=}-28(4)\,\rm MHz$ for $|v{=}35\rangle$.

To summarize, we identified promising intermediate states for two-photon coupling of Feshbach molecules to the absolute singlet ground state. In the next section we demonstrate this coupling directly. We first determine the previously uncertain location of the ground state via Autler-Townes spectroscopy and then perform coherent dark state spectroscopy for a precise measurement of the binding energy.


\section{Ground state spectroscopy of $^{23}$Na$^{40}$K}
\label{s:groundstate}
\begin{figure}
\begin{center}
  \includegraphics[width=0.8\columnwidth]{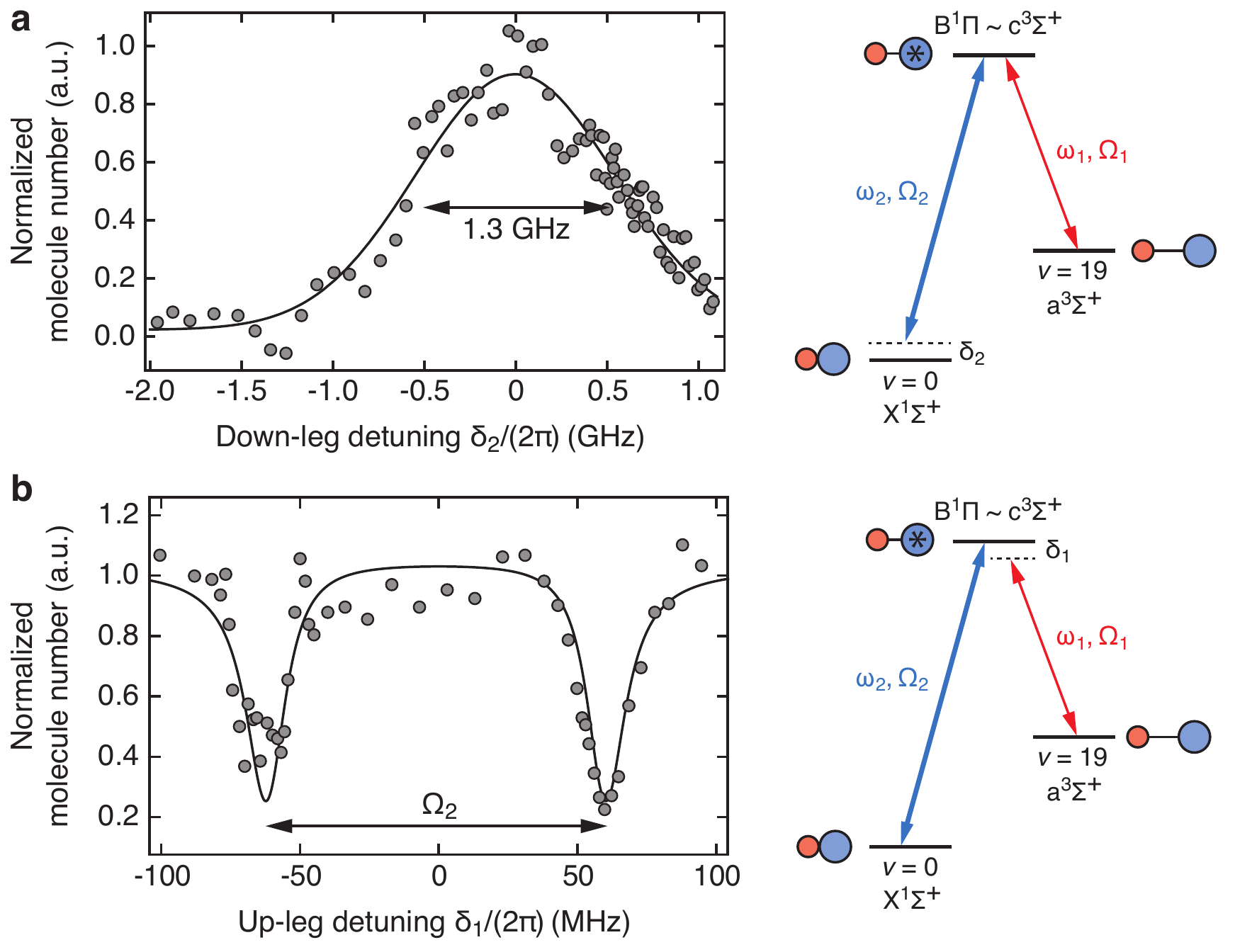}
\end{center}
  \caption{\label{fig:twophoton} Two-photon ground state spectroscopy of $^{23}$Na$^{40}$K via the $J {=}1$ level at $360.22115$ THz (compare Fig.~\ref{fig:finespectra832}). (a) The up-leg frequency $\omega_1$ is kept on resonance, while the down-leg frequency $\omega_2$ is scanned, varying the detuning $\delta_2$ (see schematic). Autler-Townes splitting of the excited state protects the Feshbach molecules from being excited, when $\omega_2$ approaches the ground state resonance. The full-width half maximum of the protection window is about 1.3~GHz. The solid line serves as a guide to the eye. (b) Autler-Townes splitting is mapped out by keeping the down-leg frequency $\omega_2$ on resonance and scanning the detuning $\delta_1$ of the up-leg frequency $\omega_1$. The line shape is reproduced with a model based on the three-level master equation (solid line). The peak distance yields a Rabi coupling of $\Omega_2 = 2 \pi \times 123(5)$ MHz.}
\end{figure}

Before our study, the latest value on the position of the absolute rovibrational ground state of bosonic $^{23}$Na$^{39}$K, relative to the dissociation limit of unbound atoms, was $5211.75(10) \,\rm cm^{-1}$~\cite{gerd08nak}. The relatively small error bar is a testament to the long history of high-resolution laser spectroscopy on NaK. For conventional heatpipe or molecular beam experiments, the main challenge is to precisely pin-point the dissociation energy, as the most weakly bound state of the triplet potential had not been directly observed.
Working with Feshbach molecules, associated from free atoms, overcomes this problem.
The measurement of interspecies Feshbach resonances between $^{23}$Na and $^{40}$K fixed the position of the weakest bound state relative to the dissociation energy~\cite{Park:2012,wu2012NaK}. Precise knowledge of the long-range van der Waals forces then allows in principle to assign absolute values to the binding energy of previously observed energy levels.

Our observation of the singlet-rich feature in the ${{\rm B}^1\Pi | v{=}4 \rangle{\sim} {\rm c}^3\Sigma^+ | v{=}25 \rangle {\sim}{\rm b}^{3}\Pi | v{=}59 \rangle}$ complex (see Fig.~\ref{fig:finespectra832}(d)) allows for a direct measurement of the singlet rovibrational ground state binding energy using two-photon spectroscopy. The dominant spin singlet character of this state results in strong coupling to the singlet ground state, which in turn induces a large Autler-Townes splitting of the excited state.

For conducting two-photon spectroscopy, similar to the experimental procedure of the previous section, we first prepare an ultracold gas of $^{23}$Na$^{40}$K Feshbach molecules in a crossed optical dipole trap. Then, the molecules are simultaneously exposed to a weak up-leg laser 1 resonantly addressing the singlet dominated $J{=}1$ state of the ${\rm B}^{1}\Pi |v{=}4\rangle {\sim} {\rm c}^{3}\Sigma^{+} | v{=}25\rangle$ branch, and a strong down-leg laser 2 coupling the ${\rm X}^{1}\Sigma^{+} | v{=}0, J{=}0\rangle$ state to this intermediate state (see Fig.~\ref{fig:cartoon}). After a set exposure time, an absorption image of the remaining Feshbach molecules is taken by imaging the $^{40}$K component of the molecules. The number of remaining Feshbach molecules is recorded as a function of the down-leg laser detuning.

In the absence of the down-leg light, the up-leg laser would simply remove most Feshbach molecules from the trap. However, when the down-leg laser is strong and sufficiently close to resonance, it shifts the excited state via the AC Stark effect by more than a linewidth, thereby ``protecting'' Feshbach molecules from decay. This gives the criterion for the width of the protection window: The AC Stark shift $\Omega_2^2/2\delta_2$ due to the down-leg coupling laser at detuning $\delta_2$ needs to exceed the linewidth $\Gamma$ of the excited state, so the detuning should not exceed $\Omega_2^2/\Gamma$. If the down-leg Rabi frequency is significantly larger than the excited state linewidth, it opens up a wide protection window, facilitating the observation of the singlet ground state. In this so-called Autler-Townes (AT) regime, the protection of Feshbach molecules originates from two-level interaction between the intermediate state and the target rovibrational ground state, and requirements on phase coherence between the two Raman lasers are relaxed.

As a starting point for the study, the binding energy of the rovibrational ground state of $^{23}$Na$^{40}$K is obtained via mass-scaling the best value for $^{23}$Na$^{39}$K. For the chosen intermediate state, the resonant down-leg frequency lies around 516 THz, corresponding to 580 nm, in the visible range. For generating this wavelength, we use a continuous wave dye laser operating with Rhodamine 6G dye. The broad tunability of a dye laser together with a large output power of $>$1W has been instrumental in locating the rovibrational ground state.

In Fig.~\ref{fig:twophoton}(a) the normalized Feshbach molecule number after exposure to the Raman lasers is plotted as a function of down-leg laser detuning. As the laser frequency is tuned close to the energy difference between the intermediate state and the rovibrational ground state, we observe a large window of Autler-Townes protection of about 1.3~GHz. This signals the observation of the absolute ground state. The width of the protection window was on the order of the uncertainty in the ground state binding energy, facilitating the observation of the ground state.

To accurately measure the Autler-Townes splitting and determine the down-leg Rabi coupling $\Omega_2$, we fix the down-leg laser to be on resonance between the intermediate state and the rovibrational ground state and scan the up-leg laser frequency. In Fig.~\ref{fig:twophoton} (b), the normalized number of remaining Feshbach molecules is plotted as a function of the up-leg detuning. From the measured splitting and known beam parameters, we obtain a normalized down-leg Rabi frequency of $\bar{\Omega}_2 = 2 \pi \times 40(10)$ kHz $\times \sqrt{I/{\rm(mW/cm^2)}}$, indicating strong coupling.

With the rovibrational ground state located, we accurately measure the binding energy of the X$^{1}\Sigma^{+} |v{=}0, J{=}0\rangle$ state via coherent dark state spectroscopy. In the limit of weak down-leg Rabi coupling, the transition probability to excite Feshbach molecules into the intermediate state is coherently canceled when the Raman lasers satisfy the two-photon resonance condition. In contrast to Autler-Townes protection, the protection of Feshbach molecules is now due to electromagnetically induced transparancy (EIT)~\cite{Fleischhauer:2005}, which relies on the relative phase coherence between the two Raman lasers.

\begin{figure}
 \begin{center}
  \includegraphics[width=0.95\columnwidth]{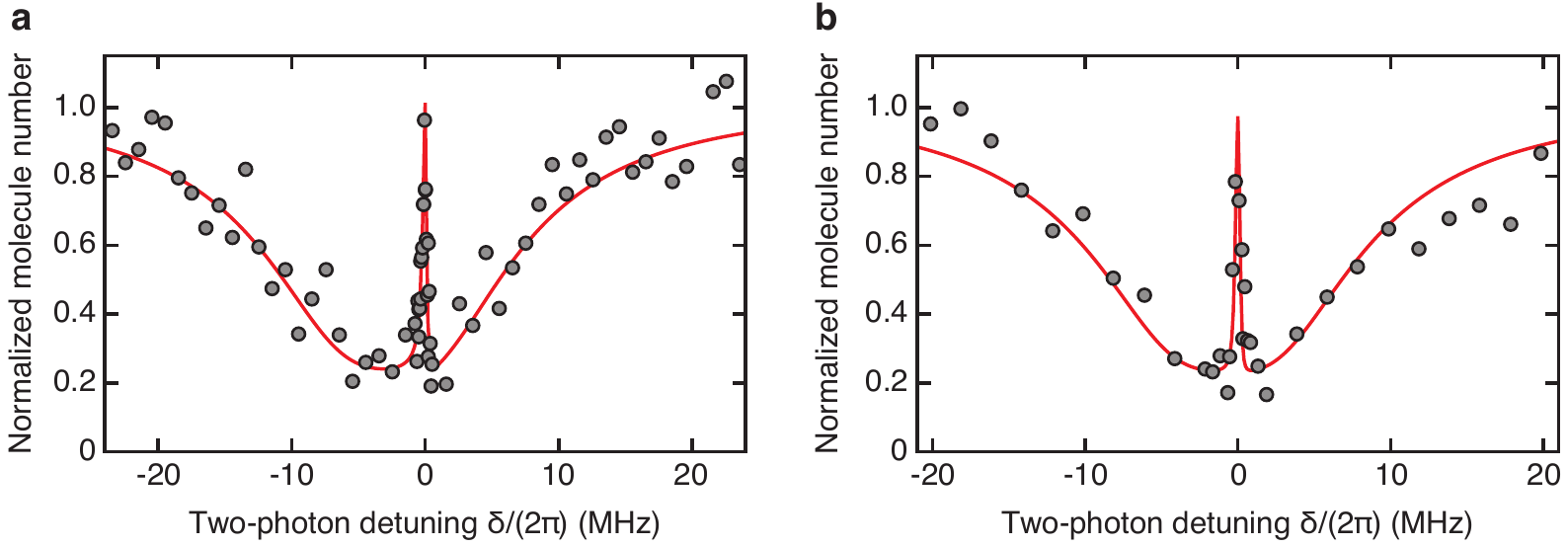}
\end{center}
  \caption{\label{fig:EIT} Observation of the absolute ground state via two-photon dark state spectroscopy. (a) Dark state spectrum via the excited state feature at 360 THz for 25~mW up-leg power (vertical polarization) and  20~$\mu$W down-leg power (horizontal polarization). The dark state feature has a width of about 200 kHz. A three-level master equation model~\cite{Bergmann:1998} matches the data for Rabi couplings $\Omega_1 = 2 \pi \times 0.20(4)$ MHz, $\Omega_2 = 2 \pi \times 2.6(5)$ MHz and a detuning $\delta_2 = 2 \pi \times 2.5(3)$ MHz. This yields normalized Rabi couplings of $\bar{\Omega}_1 = 2 \pi \times 0.25(5)$ kHz $\times \sqrt{I/{\rm(mW/cm^2)}}$ and $\bar{\Omega}_2 = 2 \pi \times 100(20)$ kHz $\times \sqrt{I/{\rm(mW/cm^2)}}$.
The linewidth of the excited state is $\Gamma_{|{\rm E}\rangle}= 2 \pi \times 10(1)$~MHz.  (b) Dark state spectrum observed via the excited state feature at 372 THz using vertical (diagonal) polarization for the up-leg (down-leg) laser. The master-equation model yields the couplings $\Omega_1 = 2 \pi \times 0.4(1)$ MHz, $\Omega_2 = 2 \pi \times 2.5(5)$ MHz and a  detuning $\delta_2 = 2 \pi \times 0.8(1)$ MHz. Normalized Rabi couplings are $\bar{\Omega}_1 = 2 \pi \times 0.45(10)$ kHz $\times \sqrt{I/{\rm(mW/cm^2)}}$ and $\bar{\Omega}_2 = 2 \pi \times 25(5)$ kHz $\times \sqrt{I/{\rm(mW/cm^2)}}$. The linewidth of the excited state is $\Gamma_{|{\rm E}\rangle}= 2 \pi \times 9(1)$ MHz.}
\end{figure}

The Raman laser coherence is maintained by locking both lasers 1 and 2 to a common ULE cavity using the Pound-Drever-Hall technique. The cavity is coated for a finesse of $\sim$35,000 in the vicinity of the up-leg and the down-leg transitions, and has a free spectral range of 1.5 GHz. For providing fast frequency feedback to the coupling dye laser, an electro-optic modulator is inserted into the laser cavity and driven by a high voltage, high bandwidth amplifier. Frequency tuning of the dye laser is achieved with an external acousto-optic modulator. As in the previous section, for addressing the up-leg transition, we use two grating stabilized diode lasers in a master-slave configuration.

In Fig.~\ref{fig:EIT}(a) and (b), we show the observed dark state resonance lineshapes, using the 360 THz singlet dominanted $J{=}1$ feature in Fig.~\ref{fig:finespectra832}(d), as well as the lower $J{=}1$ feature of the 372 THz manifold in Fig.~\ref{fig:finespectra804}(b), respectively. For detection, we record the number of remaining Feshbach molecules after exposure to the Raman lasers as a function of up-leg laser detuning. The down-leg detuning is fixed to be on resonance. The observed lineshapes are fitted by a simplified three-level master equation model, which takes the down-leg Rabi frequency as a free parameter.
For convenient absolute calibration of the down-leg frequency $\omega_2$, we employ iodine spectroscopy and reference to the iodine atlas~\cite{iodineatlas}, cross-checked with the program Iodine Spec~\cite{iodinespec5}.
The frequency difference between the intermediate state feature of Fig.~\ref{fig:finespectra832}(d) and the rovibrational ground state was measured to be $\omega_2 = 2 \pi \times 516.472613(3)$~THz, accurate to the MHz-level. The corresponding up-leg frequency $\omega_1 = 2\pi \times 360.221309(1)\,\rm THz$ was calibrated using an optical frequency comb. We obtain a binding energy of $D_{0}^{({\rm X})}$=5212.0447(1)~cm$^{-1}$ for the rovibrational singlet ground state of  $^{23}$Na$^{40}$K, relative to the dissociation threshold and the hyperfine center of mass of the constitutent atoms. Adding the zero-point energy 61.5710 cm$^{-1}$, the depth of the ${\rm X}^1\Sigma^+$ ground state potential is $D_{e}^{({\rm X})}$=5273.6157(1) cm$^{-1}$. This corresponds to a thousand-fold reduction in the uncertainty compared to previous determinations \cite{gerd08nak}.


\section{Conclusion}

In this work, we have identified a suitable pathway towards the production of ultracold ground state molecules of fermionic $^{23}$Na$^{40}$K, starting with Feshbach associated pairs of ultracold atoms. One-photon spectroscopy directly revealed the ${\rm c}^{3}\Sigma^+$ triplet state of NaK, whose levels had only been observed before in the presence of strong perturbation by nearby singlet states. We identified two strongly perturbed ${\rm c}^{3}\Sigma^+{\sim}{\rm B}^1\Pi{\sim}{\rm b}^3\Pi_0$ manifolds in $^{23}$Na$^{40}$K, which feature levels of strongly mixed singlet and triplet character. Hyperfine-resolved spectroscopy revealed in one case almost equal singlet-triplet mixing.
These excited energy levels were shown to be ideal stepping stones for a two-photon process that connects the weakly bound Feshbach molecular state to the $|v{=}0, J{=}0\rangle$ absolute rovibrational singlet ground state. We performed Autler-Townes spectroscopy to detect the absolute ground state, and demonstrated coherent two-photon dark state spectroscopy for a precision measurement of the singlet ground state binding energy. The two resonantly mixed excited state manifolds should allow an efficient stimulated rapid adiabatic passage (STIRAP) from Feshbach molecules to absolute singlet ground state molecules. In this ground state, $^{23}$Na$^{40}$K molecules are chemically stable against binary collisions \cite{zuch10mol} and possess a large induced electric dipole moment of 2.72 Debye. The results presented here pave the way towards the creation of a strongly interacting dipolar Fermi gas of chemically stable NaK molecules \cite{Park:2015:2}.

\section*{Acknowledgments}

We would like to thank Eberhard Tiemann, Thomas Bergeman, Amanda Ross, Robert Field, Rainer Blatt, Marco Prevedelli, and Jun Ye for fruitful discussions, and Cheng-Hsun Wu, Jennifer Schloss and Emilio Pace for experimental assistance. This work was supported by the NSF, AFOSR PECASE, ARO, an ARO MURI on ``High-Resolution Quantum Control of Chemical Reactions'', an AFOSR MURI on ``Exotic Phases of Matter'', and the David and Lucile Packard Foundation.

\section*{References}

\bibliographystyle{iopart-num}

\bibliography{References_forNJP_deletesort}

\end{document}